\documentclass[twocolumn,           
               showpacs,            
               preprintnumbers,     
               aps,                 
               prd,                 
               a4paper,             
               superscriptaddress,  
               nofootinbib,         
               tightenlines,        
               floats,floatfix      
               ]{revtex4}
\usepackage{graphicx}
\usepackage{latexsym}
\usepackage{amsmath,amssymb}        
\usepackage[draft=false]{hyperref}

\def\be{\begin{equation}}
\def\ee{\end{equation}}
\def\bea{\begin{eqnarray}}
\def\eea{\end{eqnarray}}

\def\R{{\cal{R}}}
\def\Rc{{\cal{R_{\rm{c}}}}}

\def\Dc{{\Delta_{\rm{th}}}}

\begin{document}


\title{A new calculation of the mass fraction of primordial black holes}
\author{Anne M. Green}
\affiliation{Astronomy Centre, University of Sussex, Brighton BN1 9QH, United
Kingdom}
\author{Andrew R.~Liddle}
\affiliation{Astronomy Centre, University of Sussex, Brighton BN1 9QH, United
Kingdom}
\author{Karim A.~Malik}
\affiliation{Physics Department, University of Lancaster, Lancaster LA1 4YB, 
United
Kingdom}
\author{Misao Sasaki}
\affiliation{Yukawa Institute for Theoretical Physics, Kyoto University, Kyoto 
606-8502, Japan}
\date{\today}
\pacs{04.70.-s, 98.80.-k \hfill astro-ph/0403181}
\preprint{astro-ph/0403181}

\begin{abstract}
We revisit the calculation of the abundance of primordial black holes
(PBHs) formed from primordial density perturbations, using a formation
criterion derived by Shibata and Sasaki which refers to a metric
perturbation variable rather than the usual density contrast. We
implement a derivation of the PBH abundance which uses peaks
theory, and compare it to the standard calculation based on a
Press--Schechter-like approach. We find that the two are in reasonable
agreement if the Press--Schechter threshold is in the range
$\Dc \simeq 0.3$ to $0.5$, but advocate use of the peaks theory expression 
which is 
based on a sounder theoretical footing.

\end{abstract}

\maketitle

\section{Introduction}

Primordial black holes (PBHs) may have formed during the early
Universe, and if so can have observational implications at the present
epoch, either from effects of their Hawking evaporation or from a
contribution to the present dark matter density \cite{carr,lims}. That
there is no unambiguous observational evidence of PBHs is a
significant constraint on some possible types of early Universe
physics.  In particular, they are the only known way of constraining
the density perturbation spectrum on extremely short scales, and
indeed until fairly recently provided the most powerful upper limit on
the spectral index of perturbations with an exactly power-law power
spectrum.

However, the abundance of PBHs formed from a given initial power
spectrum remains uncertain. The traditional calculation takes the same
form as the Press-Schechter calculation much used in large-scale
structure studies \cite{PS}, where the density field is smoothed on a
mass scale $M$ (in this application taken to be at the time of horizon
crossing), and those regions where the density contrast exceeds a
threshold value $\Dc$ are assumed to form PBHs with mass greater
than $M$. However the correct value for the threshold is quite
uncertain. The `standard' value of 1/3 for a radiation-dominated
Universe was derived by Carr \cite{carr} (see also Ref.~\cite{harr}),
but was probably only ever intended as an order-of-magnitude
estimate. Subsequently, Niemeyer and Jedamzik~\cite{nj} carried out
numerical simulations of the collapse of isolated regions and found
the threshold for PBH formation, in terms of the relative excess mass
within the horizon, to be $\Delta M / M_{{\rm h}}= 0.7$.  However,
Shibata and Sasaki~\cite{ss} have pointed out that they formulate
their initial data after horizon crossing, and hence their criterion
cannot be related to the initial perturbations produced by, for
instance, a period of inflation.

More recently, Shibata and Sasaki~\cite{ss} devised a new approach to
the formation of individual PBHs, seeking to find criteria on the
metric perturbation rather than the density field, and in a form which
can be applied to superhorizon initial perturbations. They were able
to specify a criterion in terms of whether the initial central value
of a particular metric perturbation variable $\psi$ exceeds a
threshold value. In this paper, we investigate the implications of
this result for the abundance of PBHs formed.

\section{The PBH formation criterion}

We briefly describe the PBH formation criterion of Shibata and Sasaki,
whose paper can be consulted for the full details \cite{ss}. They define a
metric variable $\psi$ from the spatial part of the metric
on uniform-expansion hypersurfaces [their
Eq.~(2.2)] as 
\be
g_{ij}=a^2\psi^4\gamma_{ij}\,,  
\ee
where $\gamma_{ij}$ is the metric of the spatial 3-sections
(throughout we assume a flat background and only consider scalar
perturbations).  Shibata and Sasaki numerically explored a range of
initial configurations, all spherically symmetric, for the metric
variable $\psi$ in a radiation-dominated Universe, and were able to
show that the central value of $\psi$, denoted $\psi_0$, was a good
indicator of PBH formation. They found that PBH formation took place
provided $\psi_0$ exceeded a threshold value $\psi_{0,\mathrm{th}}$.
The precise value of this threshold depended on the environment of the
initial configuration, and lay in the range
from 1.4 for a density peak
surrounded by a low-density region, to 1.8 for a peak surrounded
by a flat Friedmann--Robertson--Walker (FRW) region.

We wish to relate the Shibata--Sasaki threshold criterion to
quantities given in standard linear perturbation theory, where the
spatial part of the metric tensor is given by~\cite{Bardeen}
\be 
g_{ij}=a^2\left[\left(1+2\R\right)\delta_{ij}
+2\partial_i\partial_jH_{{\rm T}}\right]\,,
\ee 
where $\R$ is the curvature perturbation and $H_{{\rm T}}$ represents
the anisotropic part. The gauge-invariant curvature
perturbation on uniform-density hypersurfaces $\zeta$ is defined
as~\cite{WMLL}
\be
\label{defzeta}
\zeta=\R-H\frac{\delta\rho}{\dot\rho}\,,
\ee
with $H$, $\rho$, and $\delta\rho$ denoting the Hubble parameter,
background density, and perturbed density, respectively, which then
gives
\be
g_{ij}=a^2\left( 1+2\zeta\right)\delta_{ij}\,,  
\ee 
on superhorizon scales, where the anisotropic part $H_T$ is negligible.
Note that on large scales uniform-density hypersurfaces coincide with
uniform-expansion hypersurfaces (also known as uniform-Hubble
hypersurfaces)~\cite{Bardeen}.

A reasonable prescription for relating $\zeta$ and $\psi$ in the
quasi-linear regime is
\be
\label{zetapsi}
\exp{(2 \zeta)}=\psi^4 \,,
\ee
since by definition $\psi^4 = \exp{(2 \Delta N)}$ where $\Delta N$ is
the difference in $e$-foldings between uniform-expansion
hypersurfaces, and we can argue that the uniform-expansion and
uniform-density slices are almost equivalent even in the non-linear
regime, so that $\zeta= \Delta N$ \cite{WMLL,ST}.  Using
Eq.~(\ref{zetapsi}), we find that the threshold values of $\psi_{0}$
($\psi_{0, {{\rm th}}}=1.4$ and $1.8$) correspond to thresholds on
$\zeta$ of $\zeta_{{\rm th}}= 0.7$ and $1.2$ respectively.

\section{The PBH abundance}

The observational constraints on the fraction of the energy density of the 
Universe
in PBHs at the time they form, $\Omega_{{\rm PBH}}(M)$, can be very roughly
summarized as 
\be 
\Omega_{{\rm PBH}}(M) 
\equiv
\frac{\rho_{{\rm PBH}}}{\rho_{{\rm tot}}} \lesssim 10^{-20} \,,
\ee
on any interesting mass scale.  Detailed examination of particular
constraints can give more accurate values for the limits at particular
masses~\cite{carr,lims}, but for our present purpose we need only have an
approximate guideline. In any event, PBH formation calculations remain
uncertain enough that high-accuracy observational constraints are
unnecessary; nevertheless the production rate is normally so sensitive
to quantities we might wish to constrain, such as the density
perturbation amplitude, that useful constraints can be extracted even
from quite approximate calculations and constraints.

\subsection{Review of the standard calculation}

The traditional PBH abundance calculation 
(e.g.~Refs.~\cite{carr,abund}) refers to a quantity which in modern terminology 
would be known as the density contrast on the comoving (velocity-orthogonal)
slicing, which we denote by $\Delta$.
The density contrast is smoothed on a scale 
$R$, and the calculation
simply integrates the probability distribution $P(\Delta(R))$ over the range
of perturbation sizes
which form PBHs: $\Dc < \Delta(R) < \Delta_{{\rm cut}}$,
where the upper limit arises since very large perturbations would
correspond to a separate closed universe in the initial conditions~\cite{ch}.
In practice $P(\Delta(R))$ is such a rapidly decreasing function 
of $\Delta(R)$ above $\Dc$ that the upper cut-off is not 
important. The threshold density is taken as $\Dc > w$, where $w=p/\rho$ is
the equation of state \cite{carr}. This cannot of course be valid in the
limit $w \rightarrow 0$, but is thought to be acceptable for the
radiation-dominated case $w=1/3$ which is the main one of interest.

The smoothed density contrast $\Delta(R, {\bf x})$ is found 
by convolving the density contrast $\Delta({\bf x})$ according to
\be 
\Delta(R, {\bf x})
= V^{-1} \int W(|{\bf x}' - {\bf x} |/R) 
\Delta({\bf x}' ){\rm d}^3 x'  \,, 
\ee 
where $R$ is the smoothing scale, $W(y)$ is the window
function used for the smoothing and $V$ is the volume of the window function.  
If the initial perturbations are gaussian, this property will be inherited by 
the smoothed density perturbation so that
\be 
P(\Delta(R)) =
\frac{1}{\sqrt{2 \pi} \sigma_{\Delta}(R) } \exp{\left( -
\frac{\Delta^2(R)} {2 \sigma_{\Delta}^2(R)} \right)} \,,
\ee 
where $\sigma_{\Delta}(R)$ is the variance of $\Delta(R,{\bf x})$, 
\be
\sigma_{\Delta}^2(R) = \int_{0}^{\infty} W^{2}(kR) {\cal
P}_{\Delta}(k) \frac{{\rm d}k}{k} \,. 
\ee 
Here ${\cal P}_{\Delta}(k) \equiv (k^3 / 2 \pi^2) \langle |\Delta_{k}| 
^2\rangle$ 
is the power spectrum of $\Delta$ and $W(kR)$ is the volume-normalized
Fourier transform of the window function used to smooth $\Delta$.
It is not obvious what the
correct smoothing function to use is; a top-hat smoothing 
function has often been used
in the past~\cite{abund} 
although it is sensitive to scales well within the 
horizon, which requires careful treatment~\cite{BKP}. 
We prefer to use a gaussian window function:
\be 
W(kR) = \exp{ \left(- \frac{ k^2 R^2}{2} \right)} \,.  
\ee

On comoving hypersurfaces there is a simple relation between the density
perturbation and the curvature perturbation (e.g.~Ref.~\cite{LLBook}) 
\be
\Delta(t,k) = \frac{2(1+w)}{5+3w} \left( \frac{k}{aH} \right)^2 {\Rc}(k) \,,
\ee
where $\Rc$ is
the curvature perturbation on comoving hypersurfaces, which coincides 
with the curvature perturbation on uniform-density hypersurfaces,
Eq.~(\ref{defzeta}), on large scales. The power spectra are 
related by
\be
{\cal P}_{\Delta}(k,t) 
= \frac{4(1+w)^2}{(5+3w)^2} \left( \frac{k}{aH} \right)^4 
{\cal P}_{{\Rc}}(k) \,.
\ee
Then at horizon crossing we have
\be
\label{horcross}
{\cal P}_{\Delta}(k) 
= \frac{4(1+w)^2}{(5+3w)^2} {\cal P}_{{\Rc}}(k) \,.
\ee

The fraction of the Universe which exceeds the threshold for PBH
formation $\Delta(M) > \Dc$ when smoothed on scale $M$,
and hence will form a PBH with mass $>M$,\footnote{Throughout we
assume for simplicity that the PBH mass is equal to the horizon mass
$M_{{\rm H}}$ corresponding to the smoothing scale. This is not
strictly true (and in fact the PBH mass appears to depend on the size
and shape of the perturbations~\cite{nj,ss}); however, this
uncertainty is not important when applying PBH abundance constraints,
due to their relatively weak mass dependence.}
is given as in Press--Schechter theory by 
\bea
\label{omegaps}
\Omega_{{\rm PBH, PS}}(\Dc >M)&  =& 2 \int_{\Dc}^{\infty} 
           P(\Delta(M)) \,   {\rm d}  \Delta(M) \nonumber \\
        & = &  {\rm erfc} 
         \left( \frac{\Dc}{\sqrt{2}\sigma_\Delta(M)} \right) \,. 
\eea
In this expression we have followed the usual Press--Schechter
practice of multiplying by a factor 2, which can be thought of as
allowing for the fact that the PBH formation happens in regions which
are overdense with respect to the mean cosmological density.

For the purpose of specific calculations in this paper, we assume a power-law
primordial power spectrum ${\cal P}_{\Rc}(k) = A_{\Rc}
(k/k_{0})^{n-1}$, so that
\be
\sigma_{\Delta}^2(M) = \frac{2(1+w)^2}{(5+3w)^2}           
\frac{ A_{{\Rc}} \Gamma[(n-1)/2]}{(k_{0} R)^{n-1}} \,.
\ee
Spergel et al.~\cite{wmap1} found, from the WMAPext+2dFGRS dataset,
that $A_{\Rc} = (0.8 \pm 0.1) \times 2.95 \times 10^{-9}$ for
$k_{0}= 0.05 {\rm Mpc}^{-1}$.

\subsection{A new calculation using peaks theory}

The Shibata and Sasaki PBH formation criterion is expressed in terms
of the peak value of the fluctuation, $\psi_0$, at $t=0$
(equivalently, at some early time when the perturbation is on
superhorizon scales, since $\psi$ is constant on superhorizon
scales). Rather than Press--Schechter, it is therefore best suited to
a calculation of the mass function using the theory of peaks, as
extensively described by Bardeen et al.~\cite{bbks}. We will apply
peaks theory to the initial value of the variable $\zeta$.

After smoothing the density field on a scale $M$, the number density of peaks 
with height greater than $\nu$, where
$\nu=\zeta_{{\rm th}} / \sigma_{\zeta}(M)$, is given
(for high peaks) by \cite{dor,bbks}
\be 
n_{{\rm peaks}}(\nu, M) = \frac{1}{(2 \pi)^2 }
       \left( \frac{ \langle k^2 \rangle (M)}{3} \right)^{3/2} ( \nu^2 - 1)
       \, e^{-\nu^2/2} \,, 
\ee 
where $\langle k^2
\rangle (M)$ is the second moment, with respect to $k$, of the power
spectrum 
\be 
\langle k^2 \rangle (M) = \frac{1}{\sigma_\zeta^2(M)}
\int_{0}^{\infty} k^2 W^2(kR) {\cal P}_{\zeta}(k) \frac{{\rm d} k}{k}
\,.
\ee
For a power-law power spectrum ${\cal P}_{\zeta}(k) = 
A_{{\zeta}} (k/k_{0})^{n-1}$ (with $A_{{\zeta}} = A_{{\cal R}_c}$
since on super-horizon scales $\zeta ={\cal R}_c$) 
and a gaussian window function, we have
\be 
\langle k^2 \rangle (M) = \frac{n-1}{2 R^2} \,.  
\ee 

The number
density of peaks with height greater than $\nu$, when smoothed with a
gaussian filter on scale $M$, is then given by:
\begin{equation}
n_{{\rm peaks}}(\nu, M) 
 = \frac{1}{(2 \pi)^2 } \frac{(n-1)^{3/2}}{6^{3/2} R^3} (\nu^2 -1) \, 
e^{-\nu^2/2} \,,
\end{equation}
where
\be
\nu =  \left( \frac{2 (k_{0} R)^{n-1}}
        {A_{\zeta}\Gamma[(n-1)/2]} \right)^{1/2} \zeta_{{\rm th}}
        \,.
\ee
The number density of peaks is related to the fraction of the Universe in
peaks above the threshold by 
$\Omega_{{\rm PBH, peaks}}(\nu,>M) = n_{{\rm peaks}}(\nu, M) M / \rho$.
Here $M$ is the mass associated with the filter (which for a
gaussian window function is given by $M = \rho (2 \pi)^{3/2}
R^3$) so that 
\bea
\label{omegapk}
 \Omega_{{\rm PBH, peaks}}(\nu,>M) =   && \nonumber \\
&& \hspace*{-3.5cm}      \frac{ (n-1)^{3/2}}{ ( 2 \pi)^{1/2} 6^{3/2}}
 \left( \frac{\zeta_{{\rm th}}}{\sigma_{\zeta}(M)} \right)^2
       \exp{ \left( - \frac{\zeta_{{\rm th}}^2 }{2 \sigma_{\zeta}^2(M)}
     \right)} \,,
\eea
where
\be
 \sigma_{\zeta}(M) =
      \frac{5+3w}{2(1+w)} \sigma_{\Delta}(M)= \left( \frac{ A_{\zeta} 
       \Gamma[(n-1)/2]}{2 (k_{0} R)^{n-1}} \right)^{1/2}   \,.
\ee

\subsection{Comparison}

\begin{figure}[t!]
\includegraphics[width=8.5cm]{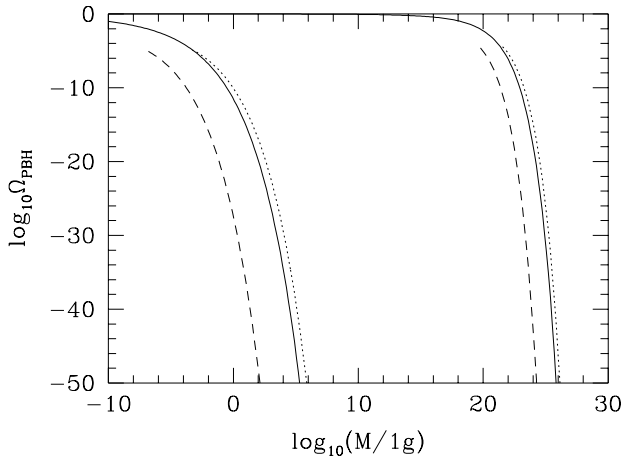}\\
\caption[f2]{\label{f2} PBH abundance as a function of horizon mass for 
power-law power spectra with $n=1.25$ and $n=1.5$ (left and right-hand 
sets of curves respectively) calculated using the Press--Schechter formalism 
with $\Dc= 1/3$ (solid 
line) and the peaks formalism with $\zeta_{{\rm th}} = 0.7$ and $1.2$ 
(dotted and
dashed respectively).}
\end{figure}

To use our results, we need to relate the comoving smoothing scale $R$ to the
horizon mass. The main case of interest is radiation domination, where $w = 
1/3$. The horizon mass is given by
\be
M_{{\rm H}} = \frac{ 4 \pi}{3} \rho (H^{-1})^{3} , 
\ee
when the scale enters the horizon, $R=(a H)^{-1}$. 
During radiation domination $aH \propto a^{-1}$, and expansion at constant 
entropy gives $\rho 
\propto g_{\star}^{-1/3} a^{-4}$ \cite{KT} (where $g_{\star}$ is the number of 
relativistic degrees of freedom, and we have approximated the temperature and 
entropy degrees of freedom as equal). This implies
\be
M_{{\rm H}} = M_{{\rm H, eq}} ( k_{{\rm eq}} R )^{2} 
                \left(\frac{g_{\star, {\rm eq}}}{g_{\star}} \right)^{1/3} \,.
\ee
In the early Universe $g_{\star}$ is expected to be of order $100$, while
$g_{\star, {\rm eq}} \approx 3$,
$k_{{\rm eq}} = 0.07  \, \Omega_{{\rm m}} h^2 \, {\rm Mpc}^{-1}$. The 
horizon mass at matter--radiation equality is given by
\be
M_{{\rm H, eq}} = \frac{ 4 \pi}{3} 2 \rho_{{\rm rad, eq}} H_{{\rm eq}}^{-3}
                = \frac{ 8 \pi}{3} 
              \frac{\rho_{{\rm rad, 0}}}{k_{{\rm eq}}^3 a_{{\rm eq}}} \,,
\ee
where $a_{{\rm eq}}^{-1} = 24 000 \Omega_{{\rm m}} h^2$
and (assuming three species of massless neutrinos) 
$\Omega_{{\rm rad, 0}} h^2= 4.17 \times 10^{-5}$ so that 
\be
M_{{\rm H,eq}}= 1.3\times 10^{49}(\Omega_{{\rm m}}h^2)^{-2} {\rm g}\,.
\ee
If we take $\Omega_{{\rm m}} h^2= 0.14$ \cite{wmapS}, then $M_{{\rm H, eq}} = 7
\times 10^{50} {\rm g}$. 

 In 
Fig.~\ref{f2} we show various calculations of the abundance
$\Omega_{{\rm PBH}}(M)$ for power-law primordial power spectra with spectral 
indices $n=1.25$ and $1.5$. The traditional calculation with $\Dc = 1/3$ is 
compared with the peaks theory calculation for the two thresholds $\zeta_{{\rm 
th}} = 0.7$ and $1.2$.  We see that the two peaks theory calculations actually 
bracket the traditional calculation. 
The high value of $\zeta_{{\rm th}}$, corresponding to the lower abundance of 
PBHs, 
is the one which corresponds to peaks surrounded by a FRW Universe, and hence 
is likely to be more appropriate for the cosmological models under discussion.

While we advocate use of the peaks theory expression Eq.~(\ref{omegapk}) to 
calculate the mass function, we see in the figure that the curves have similar 
shapes to those of the traditional calculation. In fact, if the peaks theory and 
Press--Schechter expressions were exactly the same, the thresholds would simply 
be related by Eq.~(\ref{horcross}), which in radiation domination would give 
$\Dc = 4\zeta_{{\rm th}}/9$. It turns out that this correspondence does hold 
quite accurately for our results even at the low abundances $\Omega_{{\rm PBH}} 
\sim 10^{-20}$ which are close to current observational bounds, breaking down 
only at much lower abundances where peaks theory is systematically higher than 
Press--Schechter. We therefore have quite a good correspondence: 
$\zeta_{{\rm th}}=1.2$ is equivalent to 
$\Dc \simeq 0.5$, and $\zeta_{{\rm th}}=0.7$ to $\Dc \simeq 0.3$.

\section{Discussion}

We have provided a new calculation of the abundance of PBHs generated by 
primordial density perturbations. By using a metric perturbation variable 
rather than the density contrast, a PBH formation criterion can be applied 
directly to the initial perturbation spectrum. Within this formalism, we have 
found that the PBH mass spectrum is best computed using the theory of peaks, 
rather than the standard Press--Schechter-like calculation.

Given the considerable uncertainties involved, our results do not
lead to any drastic revision of the PBH formation rate, but do put the
calculation on a sounder theoretical footing. Our mass function can be
fairly well approximated by that of the standard calculation in the
region of interest ($\Omega_{{\rm PBH}} \sim 10^{-20}$), if the
threshold density $\Dc$ is taken in the range $0.3$ to
$0.5$. This range of threshold values is however significantly lower
than the value $\Dc \simeq 0.7$ suggested by the simulations of Niemeyer 
and Jedamzik
\cite{nj}, and in fact encompasses the value $\Dc = 1/3$
used in the earliest PBH literature. However, we advocate that anyone
using our results adopts the peaks theory expression for the mass
function given by Eq.~(\ref{omegapk}).

\vspace*{10pt}

After completion of this paper Ref.~\cite{yok} was brought to our
attention. This paper uses the constraints on the metric perturbation
variable $\psi$ from Ref.~\cite{ss} to calculate the PBH abundance, but does
not use the peaks formalism.

\begin{acknowledgments}
A.M.G, A.R.L.~and K.A.M.~were supported by PPARC, and
M.S.~ by Monbukagaku-sho Grant-in-Aid for Scientific
Research (S) No.~14102004. We thank Jim Cline for asking a question
about PBH formation which led to this investigation.
\end{acknowledgments}


\end{document}